\documentclass[preprint]{aastex}
\usepackage{spr-astr-addons}
\usepackage{url}
\usepackage{rotating}

\RequirePackage{color}
\begin{document}

\title{Application of the MST clustering to the high energy $\gamma$-ray sky.\\ 
I - New possible detection of high-energy $\gamma$-ray emission associated with BL~Lac objects}

\shorttitle{High energy $\gamma$-ray emission from BL Lac objects}
\shortauthors{R. Campana et al.}

\author{R. Campana}
\affil{INAF--IASF-Bologna, via Piero Gobetti 101, I-40129, Bologna, Italy}
\and
\author{E. Massaro}
\affil{INAF--IAPS, via del Fosso del Cavaliere 100, I-00133, Roma, Italy \\ and \\ In Unam Sapientiam, Roma, Italy}
\and
\author{E. Bernieri}
\affil{INFN--Sezione di Roma Tre, via della Vasca Navale 84, I-00146 Roma, Italy}
\and
\author{Q. D'Amato}
\affil{Dipartimento di Matematica e Fisica, Universit\`a di Roma Tre, via della Vasca Navale 84, I-00146 Roma, Italy}

\abstract{
In this paper we show an application of the Minimum Spanning Tree (MST) clustering method to the 
high-energy $\gamma$-ray sky observed at energies higher than 10 GeV in 6.3 years by the 
\emph{Fermi}-Large Area Telescope. 
We report the detection of 19 new high-energy $\gamma$-ray clusters with good selection parameters
whose centroid coordinates were found matching the positions of known BL Lac objects in the 5th 
Edition of the Roma-BZCAT catalogue. 
A brief summary of the properties of these sources is presented.}


\section{Introduction}

Observational $\gamma$-ray astronomy in the GeV range is a new window on the universe opened
by the \emph{Fermi}-Large Area Telescope (LAT) mission. 
The number of photons having energies higher than a few GeV collected by this instrument
in more than six years since August 2008 is large enough to allow the detection of many Galactic and 
extragalactic sources.
The first catalogue of sources detected above 10 GeV in the first three years of the Fermi mission
\citep[1FHL,][]{ackermann13a} contains 514 objects, of which 393 are associated with Active Galactic 
Nuclei (AGNs) with a large dominance of BL Lac objects (259 sources, about 50\% of the entire catalogue).
A previous catalogue at even higher energies ($>$100 GeV) was realized by \cite{neronov11}.

In previous works \citep{campana08,campana13,bernieri13} we adapted a clustering method based on the 
Minimum Spanning Tree (hereafter MST) topometric algorithm for detecting photon concentrations in LAT 
$\gamma$-ray images of the sky, and developed some useful indicators to select clusters having 
a significance high enough to be genuine.
This method works well in not very dense photon fields with a rather uniform background and has 
a good performance in extracting weak sources appearing as quite compact clusters of four or five photons.
MST was already applied in preparing the 1FHL and the 1FGL \citep{abdo10b} and 2FGL \citep{nolan12} 
catalogues of the \emph{Fermi}-LAT collaboration.
Other clustering methods have been applied to $\gamma$-ray source searches in the \emph{Fermi}-LAT sky 
images such that by \cite{neronov11}, and DBSCAN by \cite{tramacere13}.

In this paper we show an example of MST application to the high-energy ($>$10 GeV) \emph{Fermi}-LAT sky. 
Several new rather faint clusters were found very close to the position of known 
BL Lac objects reported in the 5th Edition of the Multifrequency Blazar Catalogue Roma-BZCAT 
\citep[][hereafter 5BZCAT]{massaro14, massaro15} and having all the characteristics of genuine 
$\gamma$-ray emitters.
In the following Sections we describe the main properties of these sources which are interesting
to bridge GeV and TeV astronomy.

\section{$\gamma$-ray source detection by means of the MST algorithm}

We first present a short review our method based on MST: 
for a more complete description see \cite{campana08,campana13}.

Given a set of $N$ points (\emph{nodes}) in a 2-dimensional space, we can compute the set
$\{\lambda_i\}$ of weighted \emph{edges} connecting them. 
For a set of points in a Cartesian frame, the edges are the lines joining the nodes, weighted by
their length.
The MST is a graph without closed loops (\emph{tree}) connecting all the nodes with the minimum 
total weight, $\min [\Sigma_i \lambda_i]$. 

We divided the LAT sky in several regions to take into account the presence of the Galactic emission 
and considered the photon arrival directions as nodes in a graph, 
with the angular distance between them as the edge weight. 
Once the MST was computed for each region, a set of subtrees corresponding to clusters of photons was 
extracted by means of the following operations:
\emph{i) separation}: by removing all the edges having a length $\lambda > \Lambda_\mathrm{cut}$, the 
separation value, a set of disconnected sub-trees is obtained;
\emph{ii) elimination}: by removing all the sub-trees having a number of nodes $n \leq N_\mathrm{cut}$ 
to eliminate very small casual clusters of photons,  only the clusters having a size above this 
threshold are left. 

The separation value is usually defined in units of the mean edge length 
$\Lambda_m = (\Sigma_i \lambda_i)/N$ in the MST.
The remaining set of sub-trees provides a first list of candidate $\gamma$-ray sources and a 
\emph{secondary} selection must be done to select the most robust candidates, removing the greatest number 
of spurious/random clusters. 
\cite{campana13} proposed a useful indicator for this selection, the so-called \emph{magnitude} of the 
cluster, defined as:

\begin{equation}
M_k = n_k g_k  
\end{equation}
where $n_k$ is the number of nodes in the cluster $k$ and $g_k$ is the \emph{clustering parameter}:
\begin{equation}
g_k = \Lambda_m / \lambda_{m,k}  
\end{equation}
that is the ratio between the mean edge length over the entire tree and the mean edge length of the 
$k$-th cluster.

The magnitude $M$ quantifies the trade-off between the number of photons ($n$) and their ``clumpiness'' 
($g$): a small tree in terms of photons can be highly clustered and therefore significant.  
$M$ was shown to be a good parameter for evaluating the ``goodness'' of the accepted clusters.
On the basis of tests performed in simulated and real \textit{Fermi}-LAT fields, \cite{campana13}
 verifed that $\sqrt{M}$ has a high linear correlation with other statistical significance 
parameters, derived from a wavelet based algorithm and from maximum likelihood analysis, and that it can 
be used as a good estimator of statistical significance of MST detections.
In particular, threshold values of $M$ above 15--17 are shown \citep{campana13} to reject the 
large majority of spurious (low significance) clusters. 

Another useful parameter is the \emph{median radius} $R_m$, i.e. the radius 
of the circle centred at the cluster centroid and containing 
the 50\% of photons in the cluster, that for a cluster that can
be associated with a genuine pointlike $\gamma$-ray source should be smaller than or 
comparable to the 68\% containment radius of the instrumental Point Spread Function (PSF).
For 
\textit{Fermi}-LAT\footnote{\url{http://www.slac.stanford.edu/exp/glast/groups/canda/lat_Performance.htm}}, 
this radius varies from 0\fdg35 $\equiv$ 21\arcmin\ at 3 GeV to 0\fdg2 $\equiv$ 12\arcmin\ at 
10 GeV \citep[see also][]{ackermann13b}.
Moreover, we also expect that the angular distance between the positions of the cluster centroid 
and the possible optical counterpart are lower than the latter value.

\section{Detection of $\gamma$-ray emission associated with BL Lac objects }

LAT data (Pass 7 reprocessed v15, event class 2) above 10~GeV, covering the whole sky in the 
$\sim$6.3 years time range from the start of mission (2008 August 04) up to 2014 December 04, 
were downloaded from the FSSC archive\footnote{\url{http://fermi.gsfc.nasa.gov/ssc/data/access/}}.
Standard cuts on the zenith angle (100\degr) and data quality were applied.

The sky far from the Galactic plane ($|b|\ge20\degr$) was then divided in several broad regions to 
take into account the Galactic latitude dependence of the background level and consequently of 
the S/N ratio.
In each region the MST was computed and clusters were obtained using a cut length 
$\Lambda_\mathrm{cut} = 0.75\,\Lambda_{m}$ and a threshold number $N_\mathrm{cut}$ of 4 photons in 
a cluster.
Then a secondary selection was performed limiting the $M$ value to be higher than 15.
The large majority of selected clusters resulted in a very good positional agreement with the 
corresponding 3FGL \citep{acero15} and 1FHL sources.
As an example, on the explored extra-Galactic sky the 1FHL catalogue reports 268 sources. 
Within a matching distance equal to the 95\% containment radius (0\fdg5 $\equiv$ 30\arcmin) from 
them we found 258 corresponding MST clusters passing the primary selection. 
Almost all of these clusters ($\sim$99\%) have a magnitude $M > 15$, thus passing also the 
secondary selection.

\begin{figure}[htbp]
\centering
\includegraphics[width=8.4cm]{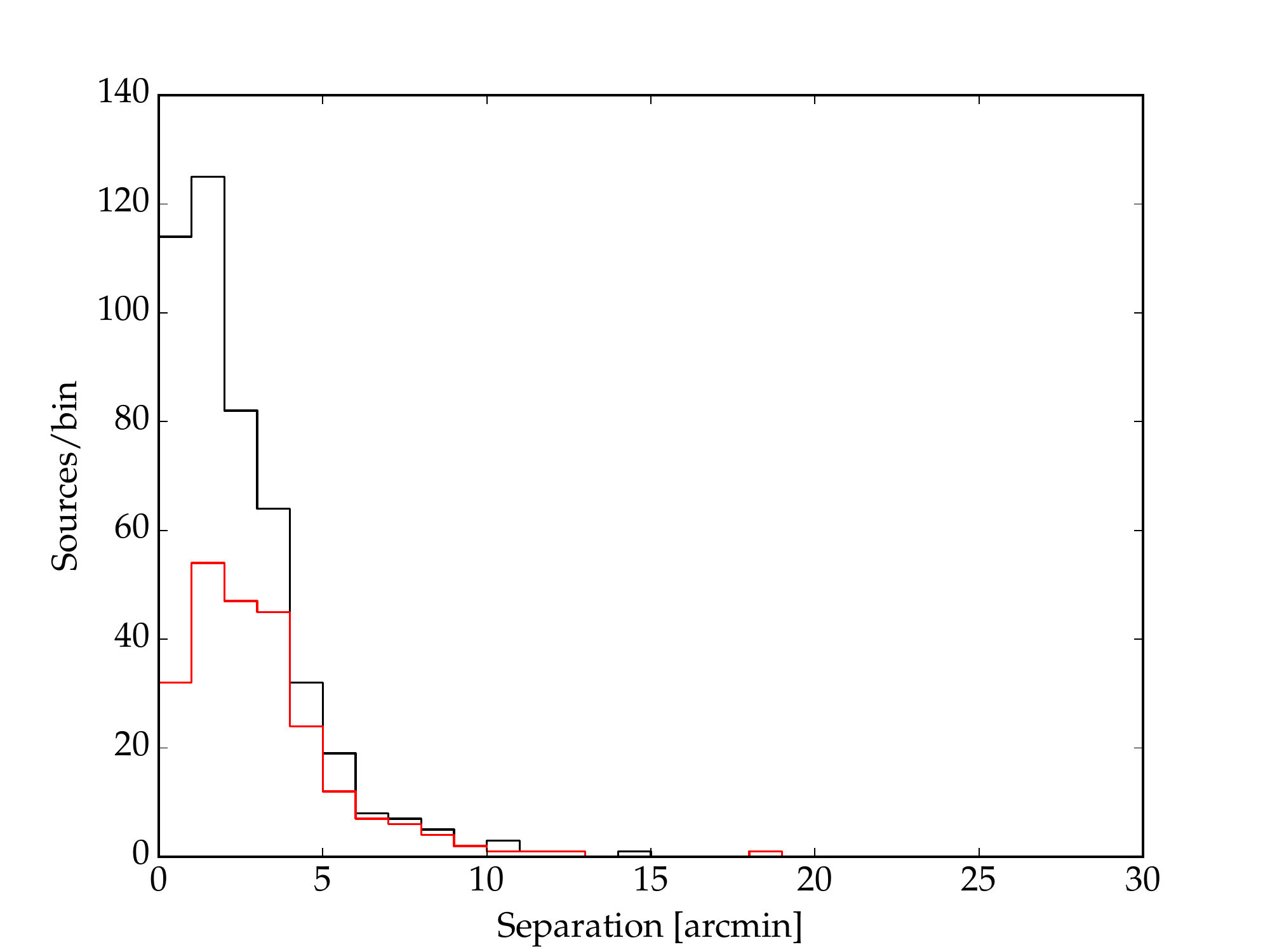}
\caption{Histogram of the angular separation between the locations of the 5BZCAT 
counterparts of the identified 3FGL sources with respect to the MST cluster centroid. 
The black line refers to all the clusters, the red line to the clusters having $n \le 12$.}
\label{hist_sep_3fgl}
\end{figure}

We also found several clusters not reported as sources in these catalogues.
We searched for associations of the unreported clusters with the 5BZCAT \citep{massaro14,massaro15},
that contains 3561 sources classified as classical BL~Lac objects (BZB), galaxy-dominated 
blazars (BZG), flat-spectrum radio quasars (BZQ) and blazars of uncertain type (BZU). 
We did not perform any \emph{a priori} selection criteria on the radio 
flux.\footnote{The fact that some sources are faint in the radio is typical of HBL objects. 
The 3LAC catalogue, for instance, includes 48 sources with a radio flux at 1.4 GHz
lower than 10 mJy (about all of HSP type). Considering the fact that we are searching for
faint gamma-ray sources (the bright ones are already in the 3FGL catalogue!) it reasonable
to assume they could be also faint in the radio band.} 
The matching radius for the association is 0\fdg1 $\equiv$ 6\arcmin, smaller than the PSF 
radius at the considered energies.
This value can be justified looking at Figure~\ref{hist_sep_3fgl}, that shows the angular 
separation between the 467 MST cluster centroids that have a correspondence with 3FGL 
sources within a 30\arcmin\ matching distance, and the safe 5BZCAT counterpart (thus with 
a very precise optical and radio location) to which the latter ones were associated. 
The large majority ($\sim$90\%) of these sources, even for weaker clusters (red histogram 
in Figure~\ref{hist_sep_3fgl}, corresponding to clusters with a number of nodes $n \le 12$) 
have a separation smaller than 6$\arcmin$. 

We found 19 high-energy photon clusters, without any correspondence with the 
3FGL and 1FHL catalogues, having thus a position very close to the accurate optical and 
radio ones of BL Lac objects.
Twelve of them have $M > 18$ and, according to the simulations by \cite{campana13} their 
probability to be spurious, based on this parameter alone, is expected to be lower than 2\%.

One must be aware that the finding of a cluster is basically an indication of the presence 
of a small region in which the photons are denser than in the surroundings, 
and it does not necessarily correspond to an actual
$\gamma$-ray source with the expected properties according to the instrumental response.
This can be estabilished only by a further analysis that takes into account these properties.
In practice, we verified in the entire extragalactic sky ($|b| > 15\degr$) that all clusters at energies above 
10 GeV having $M > 32$ and about one third of those with $15 < M < 20$ have a firm 3FGL counterpart.

Under this respect cluster analysis is important for investigating the population of sources with a 
photon number too low for obtaining a significant matching with the PSF profile.

\begin{figure*}
\centering
\includegraphics[width=0.9\textwidth]{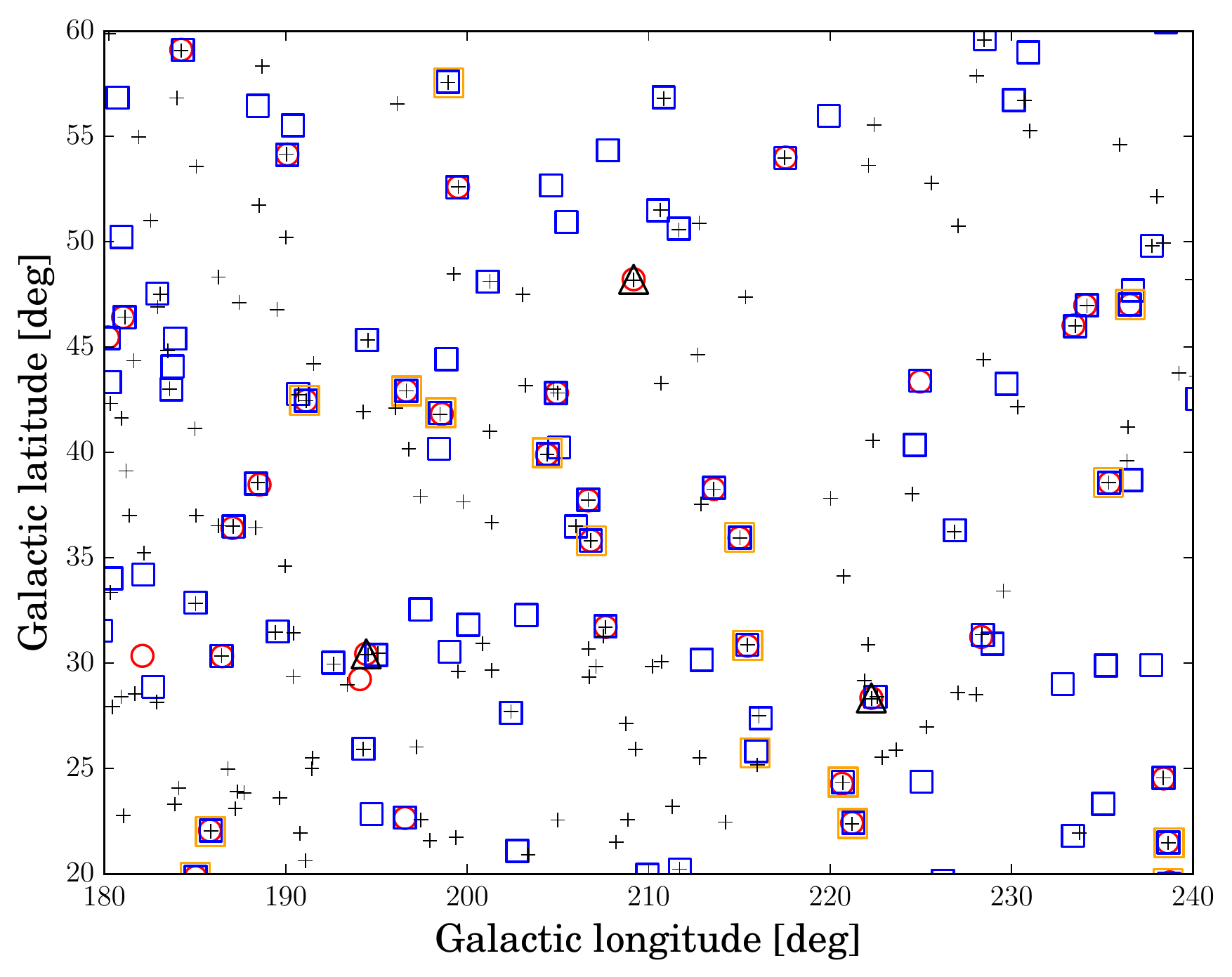}
\caption{A large region of the sky in Galactic coordinates (60\degr$\times$40\degr) showing 
the MST $\gamma$-ray source detections at energies higher than 10 GeV (red circles), superposed 
to BL Lac objects in the 5th Edition of the Roma-BZCAT (black crosses) and to the 3FGL and 1FHL 
$\gamma$-ray source catalogues (blue and orange squares, respectively). 
Note the excellent positional agreement between MST and 3FGL counterparts.
Two MST clusters are unassociated and three other correspond to BL Lac objects but not to
3FGL catalogue sources (black triangles).}
\label{skymap}
\end{figure*}

Figure \ref{skymap} presents a map of a large sky region illustrating this result: it 
contains 30 $\gamma$-ray MST clusters of which 25 are coincident with 3FGL sources (12 
also with 1FHL sources), while 3 of the 5 unassociated clusters match the position of 
BL Lac objects.

\begin{table*}
\caption{New MST sources observed at $E > 10$ GeV and their 5BZCAT associations. 
Columns 2--7 report the MST parameters of the MST cluster. 
Source data in columns 8--11 are from 5BZCAT.  
Magnitudes with the r code are from SDSS DR10; the symbols : and :: indicate uncertain and 
very uncertain data. Column 12 shows the angular separation between the cluster centroid 
and the 5BZB source. The last column reports the detection significance in equivalent Gaussian 
standard deviations, see text for details. }
\label{table1}
\scriptsize{
\begin{tabular}{lllllllllllll}
\tableline
MST source & RA      & DEC   & $n$ & $g$ & $M$ & $R_m$ &    BZCAT       & $z$    & $R$   & Log $\Phi_\textsc{xr}$  & Sep. & $P$  \\
	    & J2000   & J2000 &     &     &     &   deg & association    &        & mag   &                         &      & $\sigma$ \\
\tableline                                                        
MST~0204$-$3333 &  31.166 & -33.559 & 5  & 4.66 & 23.27 & 0.09 & 5BZB~J0204$-$3333 & 0.617  & 19.1  & 1.38  & 5\farcm7   &  4.34 \\
MST~0236$-$2940 &  39.007 & -29.671 & 9  & 2.47 & 22.20 & 0.16 & 5BZB~J0235$-$2938 & -      & 19.0  & 1.50  & 5\farcm6   &  5.38 \\
MST~0336$-$0347 &  54.095 &  -3.795 & 7  & 2.26 & 15.81 & 0.16 & 5BZB~J0336$-$0347 & 0.162: & 16.3  & 0.51  & 0\farcm3   &  5.48 \\
MST~0801$+$6442 & 120.411 &  64.705 & 9  & 3.12 & 28.10 & 0.11 & 5BZB~J0801$+$6444 & 0.20   & 18.9  & 0.75  & 4\farcm6   &  5.02 \\
MST~0803$+$2440 & 120.766 &  24.683 & 6  & 2.65 & 15.90 & 0.14 & 5BZB~J0803$+$2437 & -      & 18.8r & 0.03  & 3\farcm2   &  4.42 \\
MST~0818$+$2819 & 124.626 &  28.325 & 7  & 3.60 & 25.19 & 0.07 & 5BZB~J0818$+$2814 & 0.226  & 17.4  &-0.69  & 5\farcm5   &  5.00 \\
MST~0848$+$0503 & 132.248 &   5.064 & 12 & 3.04 & 36.47 & 0.16 & 5BZB~J0848$+$0506 & -      & 20.2  & 1.91  & 5\farcm5   &  5.02 \\
MST~0932$+$1042 & 143.208 &  10.711 & 5  & 4.80 & 23.99 & 0.08 & 5BZB~J0932$+$1042 & 0.361  & 18.0r & 0.24  & 2\farcm6   &  5.10 \\
MST~0947$+$2215 & 146.941 &  22.266 & 6  & 3.20 & 19.22 & 0.09 & 5BZB~J0947$+$2215 & -      & 19.1r & -     & 1\farcm7   &  5.43 \\
MST~0951$+$7503 & 147.913 &  75.062 & 6  & 3.03 & 18.09 & 0.08 & 5BZB~J0952$+$7502 & 0.179  & 15.6  & 1.31  & 3\farcm3   &  4.70 \\
MST~1005$+$6443 & 151.457 &  64.723 & 6  & 2.53 & 15.20 & 0.09 & 5BZB~J1006$+$6440 & -      & 18.9  & 0.15  & 4\farcm0   &  3.98 \\
MST~1137$-$1706 & 174.474 & -17.115 & 5  & 3.81 & 19.05 & 0.06 & 5BZB~J1137$-$1710 & 0.600  & 19.0  & 1.64  & 3\farcm8   &  4.66 \\
MST~1216$-$0241 & 184.031 &  -2.695 & 5  & 3.44 & 17.18 & 0.06 & 5BZB~J1216$-$0243 & 0.359  & 17.2r & 1.02  & 1\farcm7   &  4.64 \\
MST~1311$+$3951 & 197.924 &  39.865 & 6  & 2.99 & 17.95 & 0.09 & 5BZB~J1311$+$3953 & -      & 19.8r &-0.16  & 1\farcm6   &  4.98 \\
MST~1352$+$5557 & 208.235 &  55.956 & 7  & 2.27 & 15.89 & 0.11 & 5BZB~J1353$+$5600 & 0.404  & 18.4r & 1.13  & 5\farcm7   &  4.07 \\
MST~1423$+$1414 & 215.856 &  14.237 & 5  & 3.29 & 16.47 & 0.04 & 5BZB~J1423$+$1412 & 0.769::& 19.1r & -     & 1\farcm9   &  4.70 \\
MST~1423$+$3736 & 215.815 &  37.615 & 10 & 2.21 & 22.10 & 0.20 & 5BZB~J1423$+$3737 & -      & 18.5r &-0.62  & 2\farcm2   &  5.46 \\
MST~1627$+$3148 & 246.780 &  31.813 & 7  & 2.97 & 20.81 & 0.05 & 5BZB~J1627$+$3149 & -      & 19.9r & -     & 1\farcm7   &  5.55 \\
MST~2211$-$0004 & 332.819 &  -0.070 & 7  & 3.41 & 23.90 & 0.09 & 5BZB~J2211$-$0003 & 0.362  & 18.7r & 0.79  & 2\farcm4   &  5.14 \\
\hline
\end{tabular}
}
\end{table*}

Table \ref{table1} lists the main properties of the newly found clusters and their 
associated BL Lac objects.
In this table  the coordinates of cluster centroids and the relevant MST parameters 
are reported, with some astrophysical data, like optical magnitudes, redshift, and the 
Log $\Phi_\textsc{xr}$, the normalised ratio between the X-ray to the radio flux:
\begin{equation}
\Phi_\textsc{xr} = 10^2 F_X / S_{1.4} \Delta\nu  
\end{equation}
where $F_X$ is in the band 0.1--2.4 keV in units of 10$^{-12}$ erg cm$^{-2}$ s$^{-1}$ 
and $S_{1.4}$ is the radio flux density at 1.4~GHz in mJy with a bandwidth $\Delta\nu$ of 1~GHz.
This dimensionless parameter \citep{massaro12} can be used as a simple indicator if a BL Lac 
object would be classified HBL (High Energy peaked BL Lacs) or LBL (Low Energy peaked BL Lacs) 
depending if its decimal logarithm is higher or lower than 0.25. 

Note that the median radius $R_m$ is within the 68\% containment PSF radius
and that the highest angular separation is 5\farcm7, smaller than the 50\% of this radius, 
confirming the robustness of the spatial association.

Assuming a prevailing uniform random origin of the diffuse background (far from the Galactic plane) 
the significance of a cluster detection can be estimated by means of the probability to find a 
cluster having the similar $N$, $g$ and $M$ in a large set of randomly distributed photons. 
We generated 10000 simulated random fields with 50, 125 and 200 points, 
reflecting the typical number of photons in the 20 different 6\degr$\times$6\degr\ fields used 
for verifying the detection significance.
Assuming that $N_s$, $g_s$ and $M_s$ are the values of these parameters for a particular source, 
the number of random generated clusters with $N \ge N_s$, $g \ge g_s$ and $M \ge M_s$,
divided by the total number of random clusters, provides the upper limit to the probability that 
the source is of spurious origin.
Moreover, one has to evaluate the probability that the found cluster is located at a short angular 
distance from a BL Lac object not already associated with a $\gamma$-ray source. 
Simple estimates for these probabilities can be obtained as the ratio between the solid angle of 
a circle having the radius equal to the angular separation between the BL Lac objects and the cluster 
centroids, to the solid angle covered by the 6\degr$\times$6\degr\ regions used for verifying the 
detection significance, eventually multiplied for the number of known BL Lac objects inside the field,
with the exclusion of those already associated with known $\gamma$-ray sources.
These ratios range from $6.2\cdot10^{-5}$ for 5BZB~J1311$+$3953 to $7.9\cdot10^{-4}$ for 
5BZB~J0204$-$3333, the source with the largest angular separation. 
Combining these two independent probabilities we obtain the final significance of a detection, 
expressed as the probability $P$  in equivalent Gaussian standard deviations, 
for a chance coincidence of a spurious cluster with a known BL Lac object (Table~\ref{table1}).
For all clusters these combined significances are above $3.98\sigma$, while for more than half of 
them it is in excess of 5$\sigma$.

Fifteen of the 19 sources reported in Table \ref{table1} do not have another 5BZCAT source associated 
within a 0\fdg5 $\equiv$ 30\arcmin\  radius. 
For three of the remaining 4 sources, the other possible associations are much weaker than the 
proposed ones, being sufficiently far away (more than 15\arcmin\ distant). 
The last source, and some of the other associations between 5BZCAT blazars and MST clusters, will 
be discussed in the following Section.
Regarding the possibility to have a counterpart to these clusters different from a BL~Lac object, 
in general in the same search radius there are only few weak radio sources (e.g. reported by the 
NVSS\footnote{\url{http://www.cv.nrao.edu/nvss/}} and 
FIRST\footnote{\url{http://sundog.stsci.edu/first/catalogs/readme.html}} catalogues) usually without a safe
optical counterpart, or for MST~1423+1414 a faint optical (SDSS) quasar undetected in the radio band.
Other possible associations as narrow-line Seyfert~1 objects or classified radio galaxies are not found.
The association with 5BZCAT objects appears therefore much more robust than for other sources.

To increase the confidence on these detections we performed a MST cluster search in rather small regions centered 
at the cluster position and having a size of about $6\degr\times6\degr$.
All clusters upon consideration were confirmed and they were the only unassociated sources with $M > 15$. The only 
exception was a cluster in the field of MST~1423$+$3736 at about 2\fdg4.
Furthermore, we performed a similar check considering also photon energies higher than 3 GeV and
again the candidate selections were found, with similar MST cluster parameters, with the only exception of 
MST~0951$+$7503, that was undetected.

We extracted also aperture photometry light curves for all the 19 candidate sources in various energy bands. With the exception of 
MST~1423$+$3736 (see Sect.~\ref{s:flaring}) there is no evidence of variable activity or flares for the other sources.

\subsection{Results from likelihood analysis}

We performed also a standard unbinned likelihood analysis for each MST cluster. 
A Region of Interest (ROI) of 10\degr\ radius was selected centered at the MST cluster 
centroid, and standard screening criteria were applied to the \emph{Fermi}-LAT data 
above 3 GeV. 
The likelihood analysis was performed considering all the 3FGL sources within 20\degr\ 
from the cluster centroid, as well as the Galactic and extragalactic diffuse emission. 
A further source with a power-law spectral distribution was assumed at the MST coordinates. 
The normalization and spectral index of all the 3FGL sources within the ROI was allowed 
to vary in the fitting, while the parameters of the sources between 10\degr\ and 30\degr\ 
from the center of the field of view were fixed to their catalogue values.
From this analysis, we derived the likelihood Test Statistics ($TS$) and fluxes in the two 
3--300 and 10--300 GeV bands. 
The results are reported in Table~\ref{table2}. 
Eight of the 19 MST clusters are significant from a likelihood analysis standpoint ($TS>25$) and 
therefore can be considered confirmed $\gamma$-ray sources; other six clusters have a ${TS}$ value
between 16 and 25, while only four have ${TS}<16$.
For MST~0848$+$0503 no significant result was obtained because of a confusion problem with a
another close blazar (see the discussion on Section \ref{disc0848}).
Fluxes above 3 and 10 GeV between a few 10$^{-10}$--10$^{-11}$ ph\,cm$^{-2}$\,s$^{-1}$ are
obtained for the safe sources whereas for those having a $TS$ value below the conventional 
significance threshold ($TS = 25$, corresponding to 5$\sigma$) upper limits were computed.
Note that three of the four clusters with the lowest significance have also low $M$ values
(15.81, 15.90, 17.95) while only MST~0801$+$6442 has a magnitude of 28.10, that is due
to the relatively high number of photons in the cluster.
This particular case will be discussed in more detail in Sect. \ref{disc0801}.

\begin{table}[htb]
\caption{Standard unbinned likelihood analysis of the \emph{Fermi}-LAT data, see text for details. 
The last two columns report photon fluxes in units of 10$^{-11}$ ph\,cm$^{-2}$\,s$^{-1}$.  
For sources below the usual significance threshold ($TS=25$) only upper limits are given.}
\label{table2}
\small{
\begin{tabular}{lrcc}
\tableline
MST cluster &  $\sqrt{TS}$  &    Flux      &    Flux    \\
            &               &  3--300 GeV  & 10--300 GeV \\
\tableline                                                        
MST~0204$-$3333 &  4.0 & $\le 6.7 $ 		& $\le 1.0 $ 		\\
MST~0236$-$2940 &  4.6 & $\le 7.2 $ 		& $\le 1.3 $ 		\\
MST~0336$-$0347 &  3.9 & $\le 5.5 $ 		& $\le 1.4 $ 		\\
MST~0801$+$6442 &  3.3 & $\le 4.0 $ 		& $\le 0.4 $ 		\\
MST~0803$+$2440 &  3.5 & $\le 4.1 $ 		& $\le 0.8 $ 		\\
MST~0818$+$2819 &  5.0 & $6.9 \pm 2.4 $		& $2.3 \pm 1.1 $	\\
MST~0848$+$0503 &  --- & 	---			& 	---		\\
MST~0932$+$1042 &  5.4 & $7.9 \pm 2.7 $		& $2.1 \pm 1.1 $	\\
MST~0947$+$2215 &  5.8 & $9.5 \pm 2.9 $		& $2.2 \pm 1.1 $	\\
MST~0951$+$7503 &  4.6 & $\le 4.2 $		& $\le 1.6 $		\\
MST~1005$+$6443 &  5.0 & $6.3 \pm 2.1 $		& $1.5 \pm 0.8 $	\\
MST~1137$-$1706 &  6.7 & $14.0 \pm 3.4 $   	& $2.5 \pm 1.1 $	\\
MST~1216$-$0241 &  5.0 & $6.4 \pm 2.5 $		& $2.4 \pm 1.2 $	\\
MST~1311$+$3951 &  2.9 & $\le 3.6 $		& $\le 2.5 $		\\
MST~1352$+$5557 &  4.8 & $\le 7.9 $		& $\le 4.8 $		\\
MST~1423$+$1414 &  4.4 & $\le 5.2 $		& $\le 1.4 $		\\
MST~1423$+$3736 & 12.1 & $22.8 \pm 3.7$ 	       & $4.3 \pm 1.5$ 	\\
MST~1627$+$3148 &  7.0 & $10.3 \pm 2.9$ 	       & $2.4 \pm 1.2$ 	\\
MST~2211$-$0004 &  4.6 & $\le 7.6 $ 		& $\le 1.6 $ 		\\
\hline
\end{tabular}
}
\end{table}

\begin{figure}[h]
\centering
\includegraphics[height=8cm]{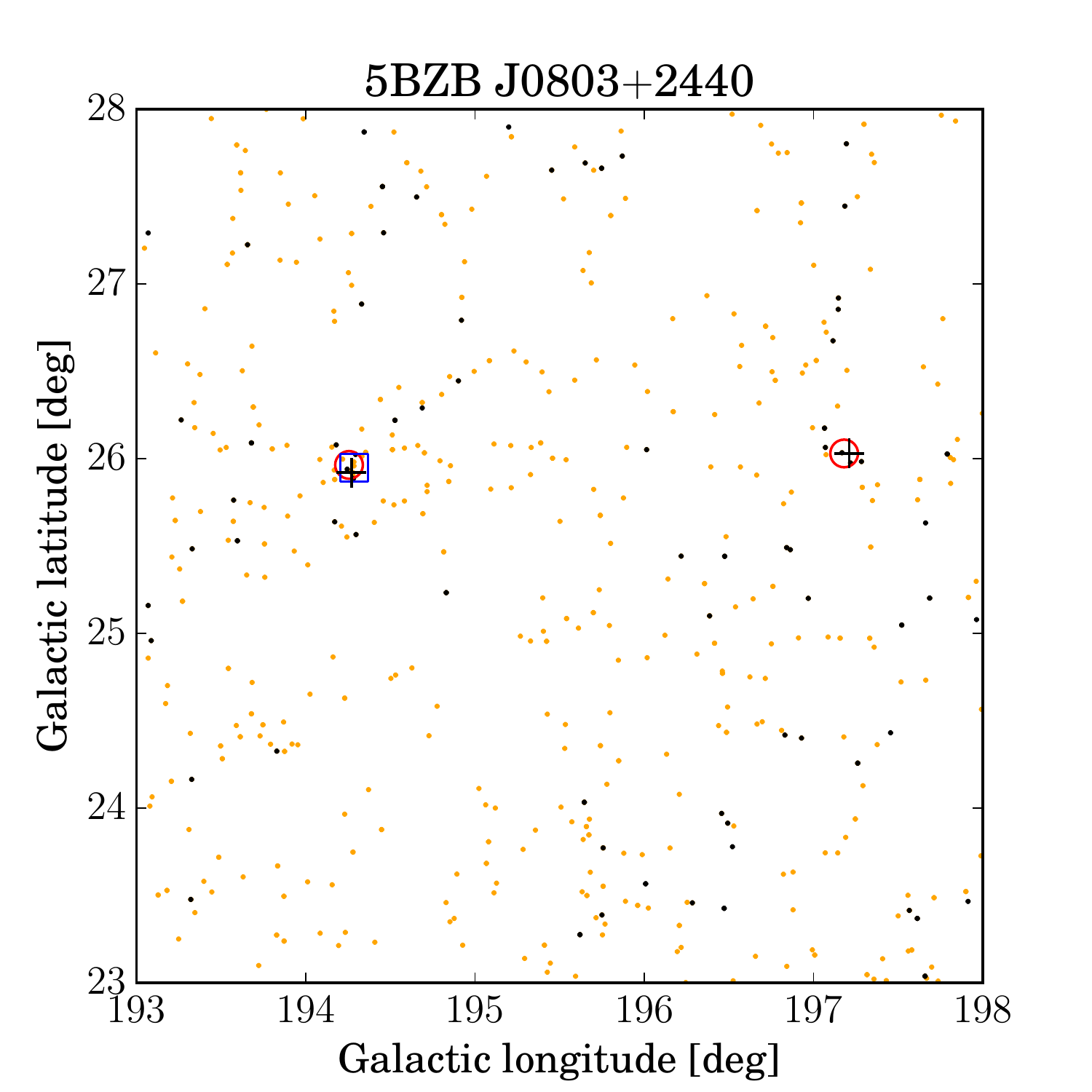}
\caption{$\gamma$-ray sky maps in Galactic coordinates of the regions centered around the 
BL Lac objects 5BZB~J0803$+$2440. 
Black circles mark the photon coordinates at energies higher than 10 GeV and orange ones
above 3 GeV;
black crosses corresponds to the optical coordinates of BL Lac objects in the 5BZCAT, blue 
open squares are the 3FGL sources, the open red circles are the MST cluster positions in 
the 10 GeV field.
}
\label{BZB_uno}
\end{figure}

\begin{figure*}
\centering
\includegraphics[height=8cm]{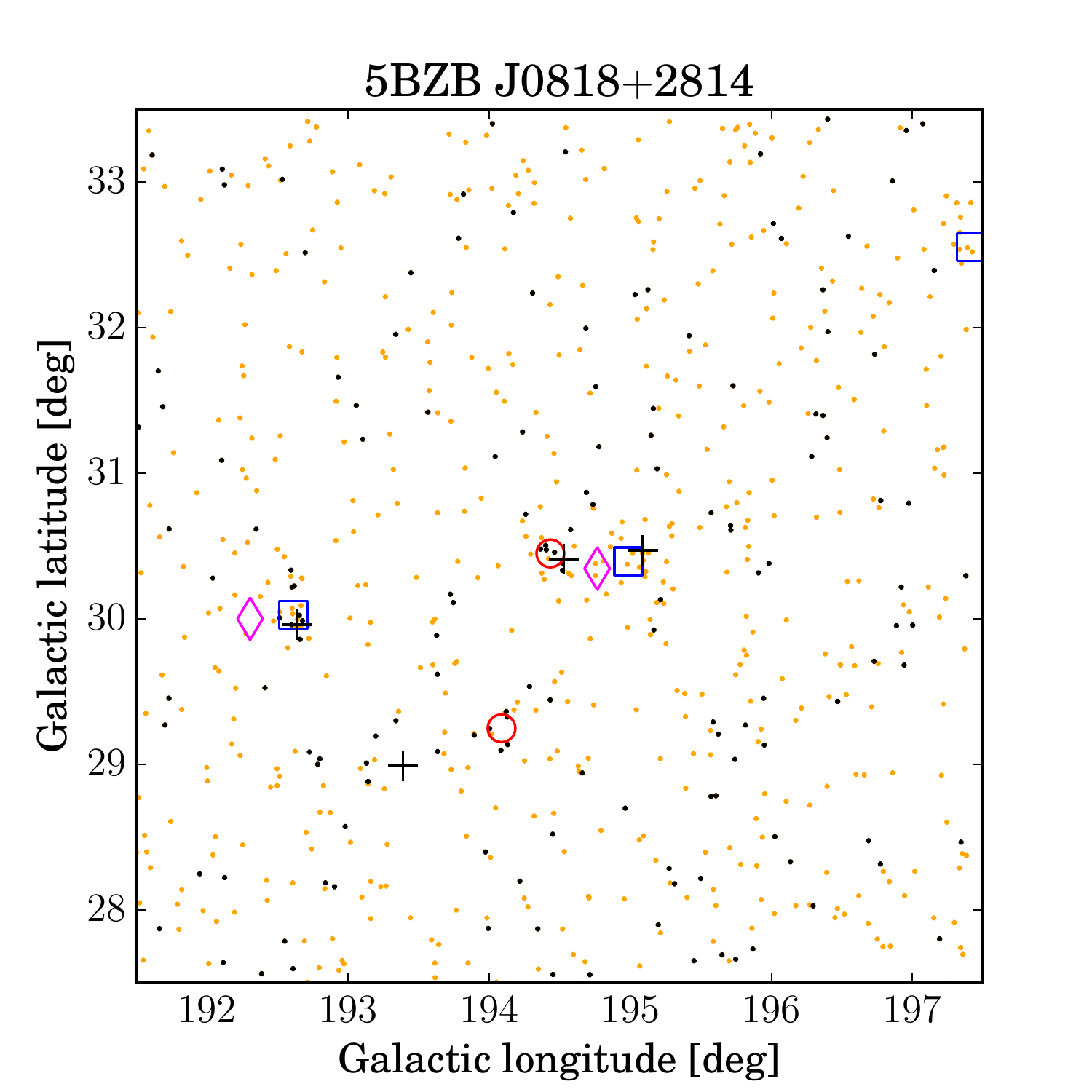}
\includegraphics[height=8cm]{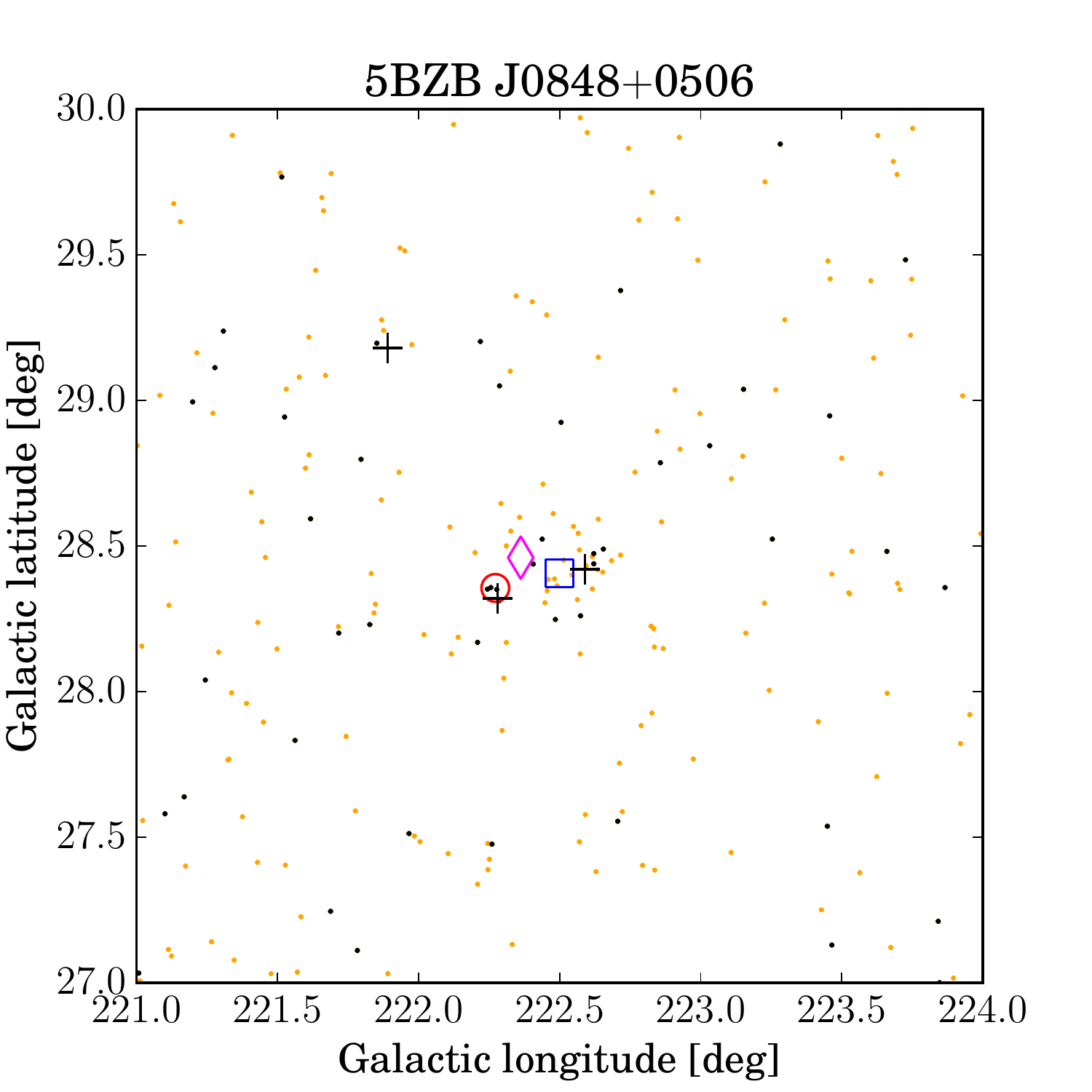}
\caption{$\gamma$-ray sky maps in Galactic coordinates of the regions centered around the 
BL Lac objects 5BZB~J0818$+$2814 (left panel) and 5BZB~J0848$+$0506 (right panel), both having 
a rather close BL Lac object. 
Orange circles mark the photon coordinates at energies higher than 3 GeV 
and the black ones at energies higher than 10 GeV;
black crosses corresponds to the optical coordinates of BL Lac objects in the 5BZCAT, blue 
open squares are the 3FGL sources, the open red circles are the MST cluster positions in 
the 10 GeV field, while the magenta diamonds are the positions of sources in the D$^3$PO catalogue.
}
\label{BZB_due}
\end{figure*}

\begin{figure*}
\centering
\includegraphics[width=8cm]{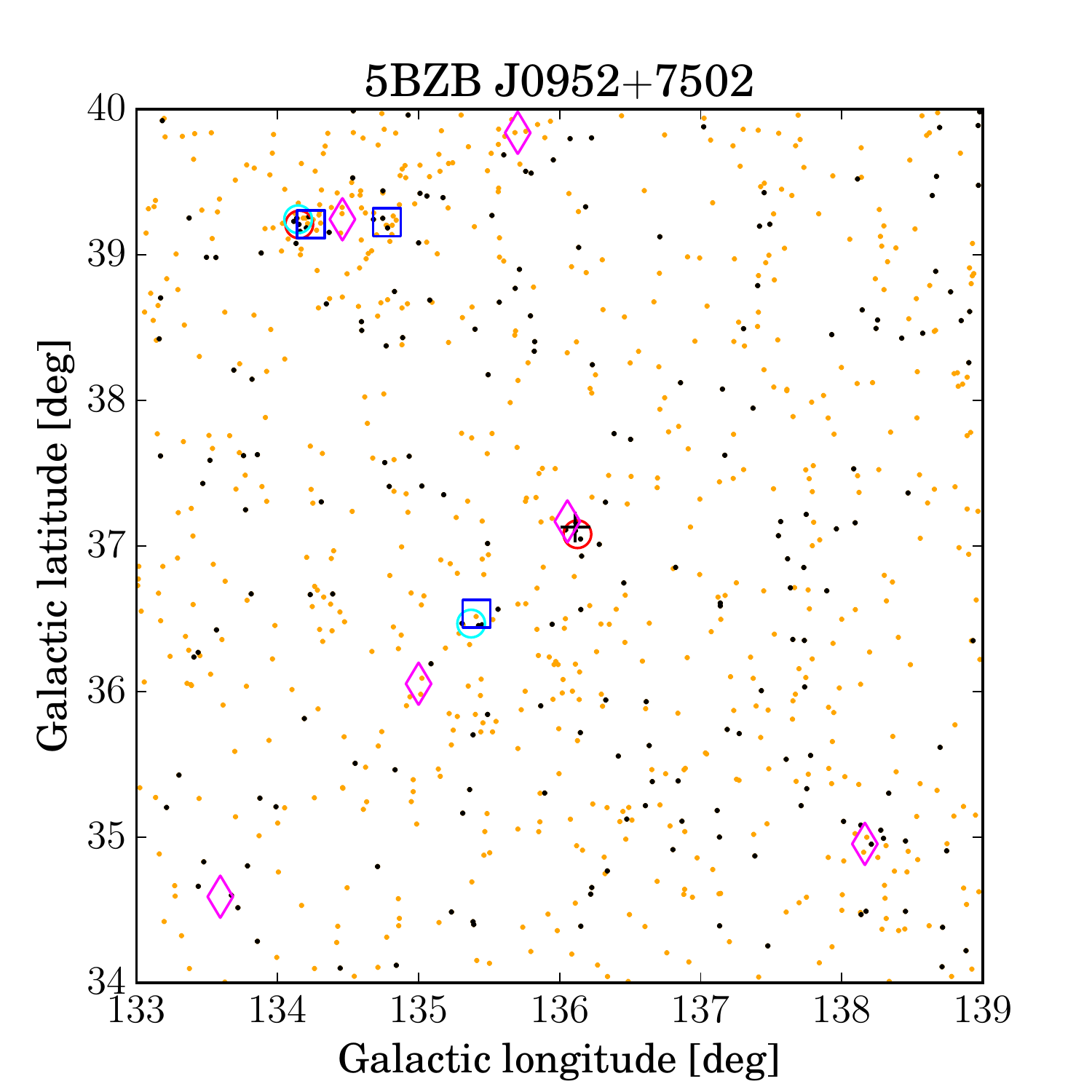}
\includegraphics[width=8cm]{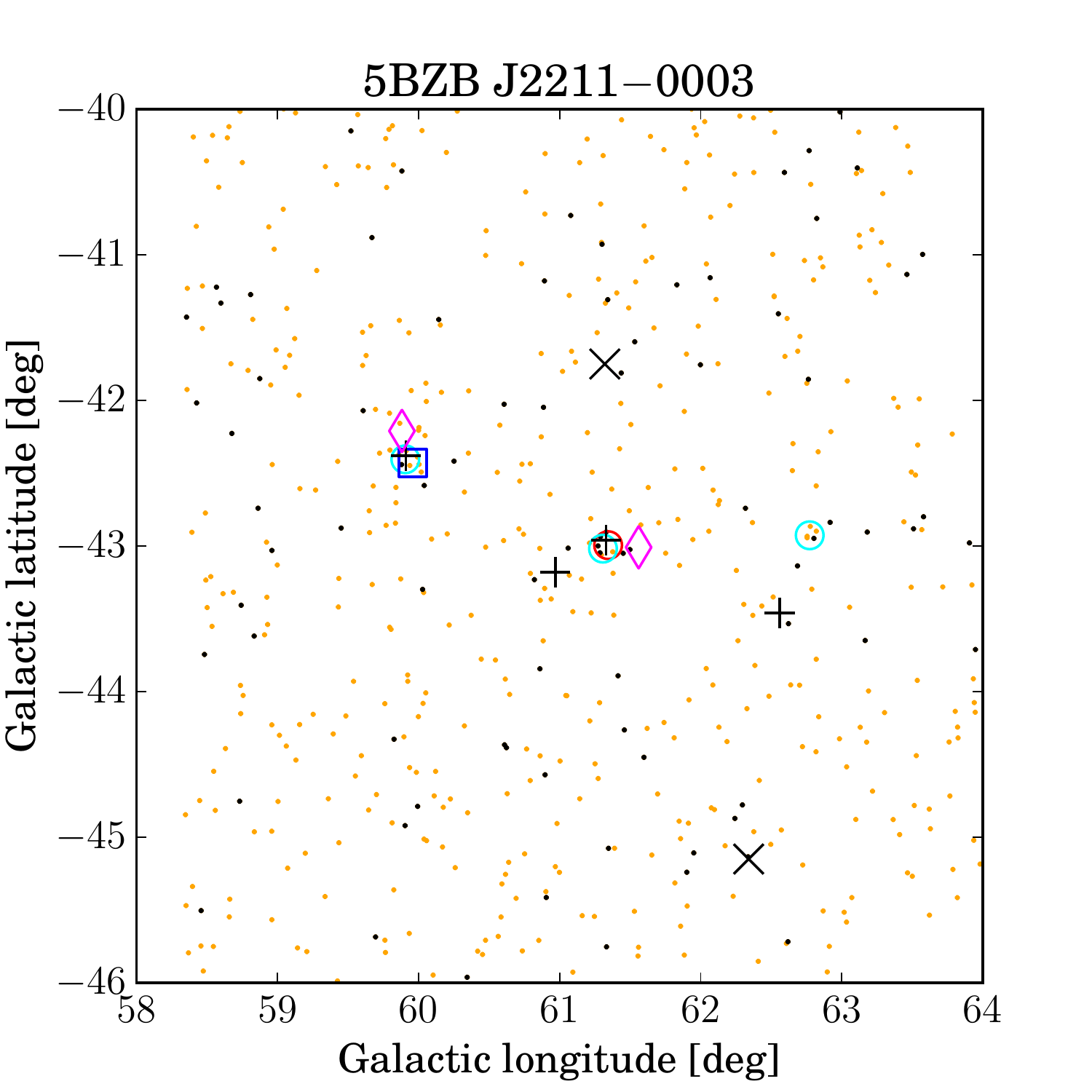}
\caption{Photon sky maps in Galactic coordinates of the regions around the BL Lac objects 
5BZB~J0952$+$7502 (left panel) and BZB J2211$-$0003 (right panel). 
The longitude scale has been multiplied by the cosine of the mean latitude for a better imaging
of angular distances.
Orange circles mark the photon coordinates at energies higher than 3 GeV, magenta circles
at energies higher than 6 GeV and the black ones at energies higher than 10 GeV;
black crosses correspond to the optical coordinates of BL Lac objects in the 5Roma-BZCAT while 
black x-symbols are blazars of other type, blue open squares are the 3FGL sources, the open red circles are 
the MST cluster positions in the 10 GeV field and the open turquoise circle at energies above 3 GeV, 
while the magenta diamonds are the positions of sources in the D$^3$PO catalogue.}
\label{BZB_tre}
\end{figure*}

\begin{figure*}
\centering
\includegraphics[height=7.95cm]{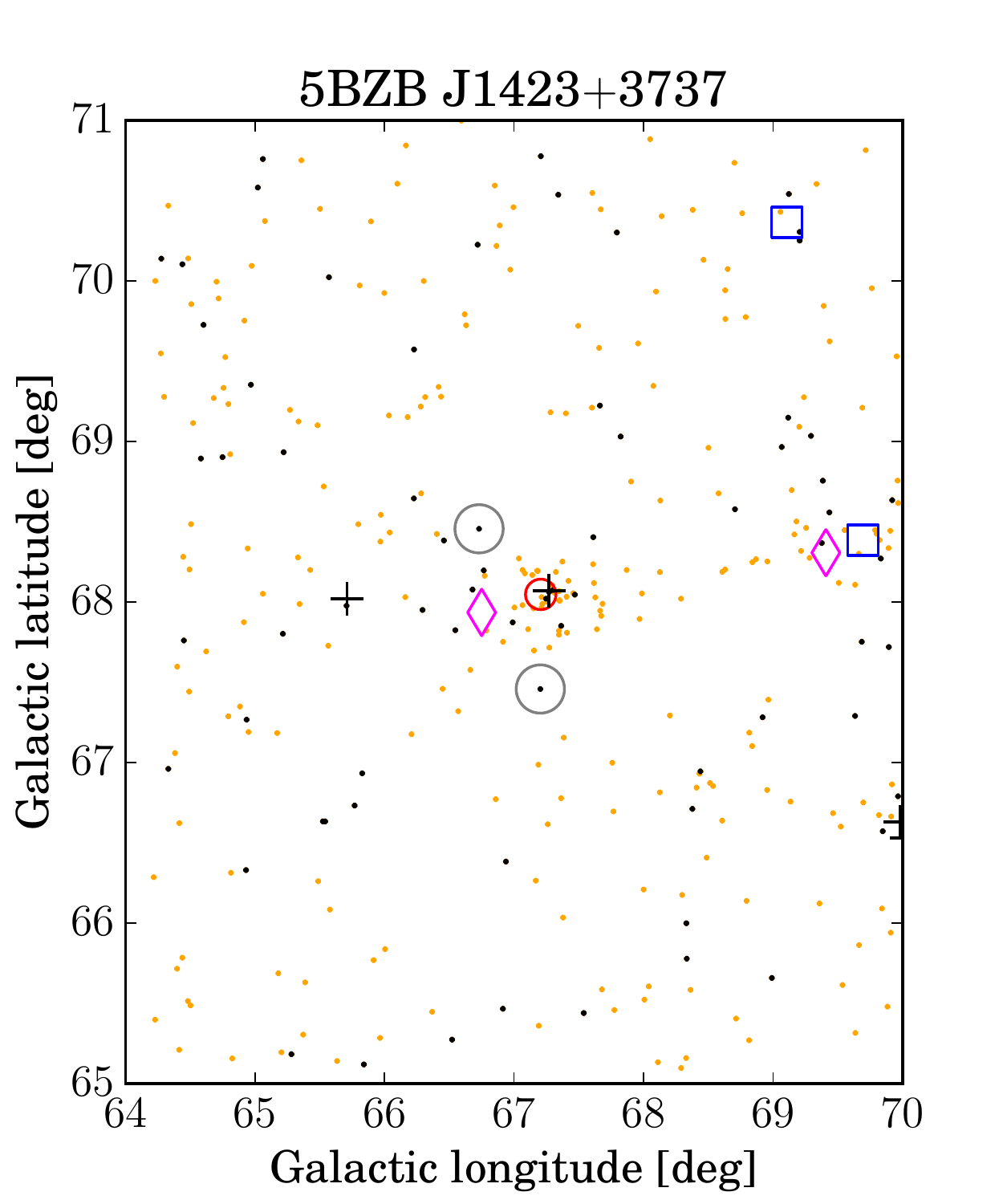}
\includegraphics[height=7.95cm]{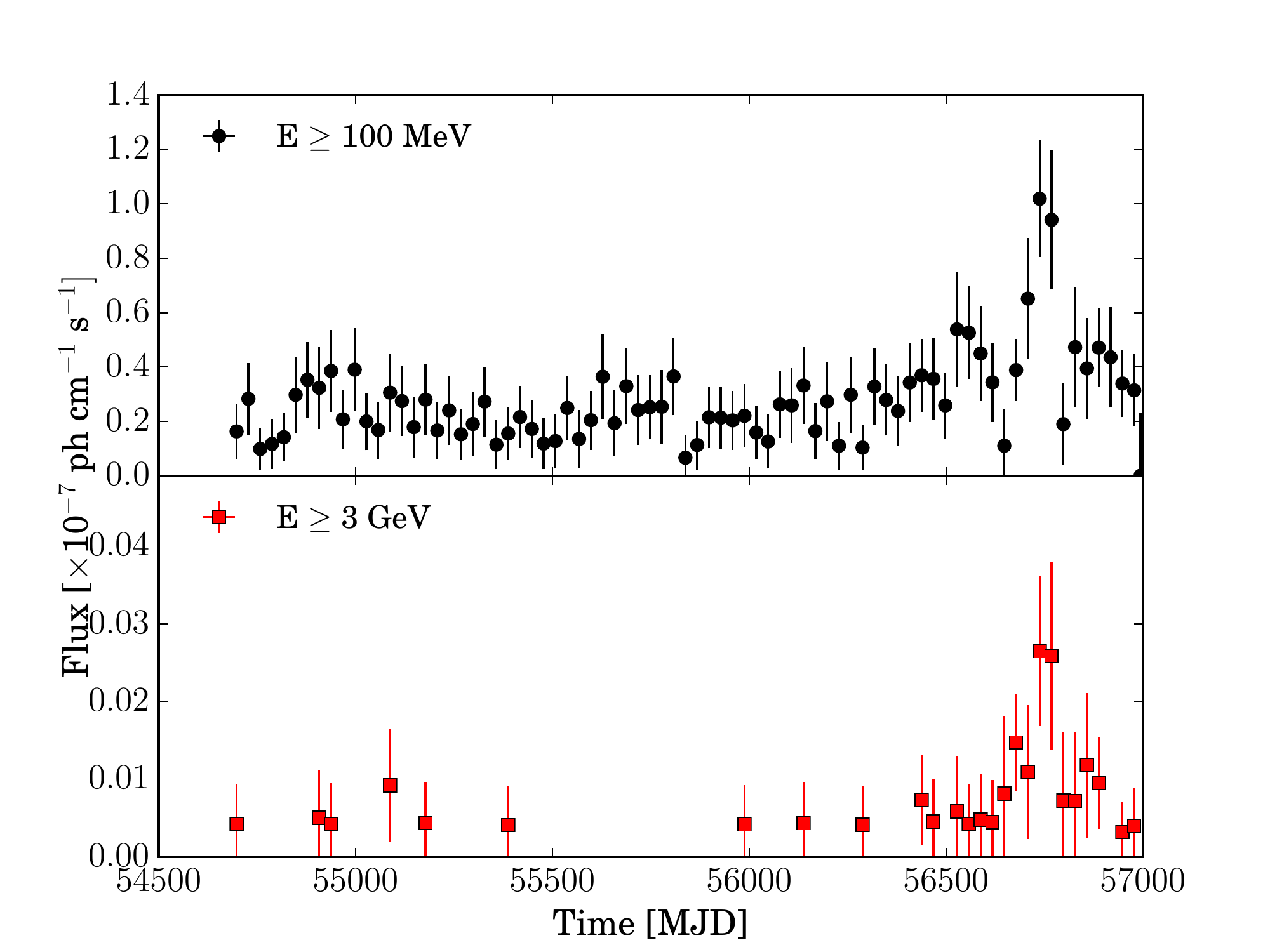}
\caption{Photon sky map in Galactic coordinates of the region around the BL Lac objects 
5BZB~1437$+$3737 (left panel), and aperture photometry light curve (right panel). 
The longitude scale has been multiplied by the cosine of the mean latitude for a better imaging
of angular distances.
Orange circles mark the photon coordinates at energies higher than 3 GeV, magenta circles
at energies higher than 6 GeV and the black ones at energies higher than 10 GeV;
black crosses correspond to the optical coordinates of BL Lac objects in the 5Roma-BZCAT while 
black x-symbols are blazars of other type, blue open squares are the 3FGL sources, the open red circles are 
the MST cluster positions in the 10 GeV field and the open turquoise circle at energies above 3 GeV, 
while the magenta diamonds are the positions of sources in the D$^3$PO catalogue.
The two gray circles indicate the locations of the two galaxy
clusters A 1902 and A 1914.}
\label{BZB_quattro}
\end{figure*}

\section{Comments on individual sources}

Many of the newly associated BL Lac objects (10 over 19) have a value of Log $\Phi_\textsc{xr}$ 
typical of HBL.
SDSS DR10 photometric data are available for four of the other six sources with an LBL value and
their $u - r$ colours are smaller than 1.4, that according to Massaro et al. (2012) can be taken
as  the discriminating value between BL Lacs having a nucleus or galaxy dominated emission.

In the following we discuss in more detail some clustering properties of a few sources of the sample.

\subsection{5BZB~J0204$-$3333}
The association of the MST photon cluster MST 0204$-$3333 with 5BZB~J0204$-$3333 should be 
considered not safely confirmed, since at a separation of 5\arcmin\ from the cluster centroid the 
rather bright flat spectrum radio source PKS~J0204$-$3328 (AT20G J020428$-$332849) is present,
for which an indication for a galaxy type counterpart of magnitude $B=23.1$ is reported \citep{titov09}.
MST analysis in the same field at energies higher than 5 GeV gives a more conspicuous cluster 
($N = 10$, $M = 31.9$) but fails to solve the positional ambiguity because photons are distributed 
around both sources.

\subsection{5BZB~J0801$+$6444}\label{disc0801}

As already noticed above the association of the cluster MST 0801$6$6442 with BZB~J0801$+$6444 appears 
somewhat unsafe because the likelyhood analysis gives a low $\sqrt{TS} = 3.3$, despite the high 
$M$ value.
A further analysis with the lower $\Lambda_\mathrm{cut} = 0.7$ reduces the number of photons in the 
cluster from 9 to 6 but increases the clustering factor $g$ to 3.50 giving again a rather robust  
$M = 20.99$. Moreover, the centroid of the new cluster is at an angular seperation from the possible
counterpart of only 2\farcm4, reducing the chance association probability. 
This result could be an indication that the cluster can be affected by some photons of the local background
that is reduced using a stronger cut length.
MST search at energies higher than 3 GeV in a $10\degr\times10\degr$ region gives a more robust detection
of a 15 photon cluster with $M = 38.00$ and centroid coordinates at a very small separation (1\farcm3) from the
BL Lac object.

\subsection{5BZB~J0803$+$2440}

This blazar and 5BZB~J1352$+$5557 are associated with the two clusters having the lowest $M$ values in the sample, 
but while the latter appears nearly validated by the ML analysis, the results on a possible $\gamma$-ray
source close to 5BZB~J0803$+$2440 give a rather poor significance (see Table 2).
Nevertheless, there is some indication for a safer detection and a more robust association of
the cluster with 5BZB~J0803$+$2440.
When selecting a smaller region of about $5\degr\times5\degr$ and applying a stronger cut with
$\Lambda_\mathrm{cut} = 0.5$ in order to reduce possbile the background contribution from distant
photons we found above 10 GeV only two clusters (see Figure~\ref{BZB_uno}): one with 5 photons and $M = 18.93$ 
nearly coincident with the previous one (RA = 120.760, DEC = 24.567) at the closer angular distance 
of 1\farcm96 from the blazar; and another cluster with only 4 photons and a lower clustering parameter
($M = 13.92$) that corresponds to the 3FGL J0758.9$+$2705 source which is very close to the other blazar 5BZG~J0758$+$2705.
When considering energies higher than 3 GeV the latter cluster is much richer and has a very
significant $M = 60.9$, while MST~0803$+$2440 is found with a very low $M$ because it disappears in 
the stronger background.
This could be an explanation of why at these energies ML results are poor.
It is therefore possible that 5BZB~J0803$+$2440 has a quite hard $\gamma$-ray spectrum that makes 
it detectable only at high energies and for this reason one can consider the association of the 
cluster with the BL Lac object as an interesting indication.

\subsection{5BZB~J0818$+$2814}

This BL Lac object is located at angular distance of about 29\arcmin\ to  
5BZB~J0819$+$2747, which is associated with the known $\gamma$-ray source 3FGL J0818.8$+$2751.
The sky map in Figure~\ref{BZB_due} (left panel) clearly shows a photon cluster around 
5BZB~J0819$+$2747 at energies higher than 3 GeV. Photons above 10 GeV are densely 
located only around the position of 5BZB~J0818$+$2814, while none is found close to the position 
of the other source.
Consider also that the cluster has a quite high $g$ and a corresponding median radius of about 
only 4\arcmin, that corresponds to about 15\% of the separation between sources.
It appears, therefore, unlikely that photons in the MST cluster could be originated by a distant 
source without any concentration at the expected position.

A source at an intermediate position between our source and that of 3FGL~J0818.8$+$2751 (see Figure~\ref{BZB_due}) 
is also reported in the D$^3$PO catalogue\footnote{The Denoised, Deconvolved, 
and Decomposed Fermi $\gamma$-ray Sky, \url{http://www.mpa-garching.mpg.de/ift/fermi/}} \citep{selig14}, 
without indication of a possibile counterpart.

\subsection{5BZB~J0848$+$0506}\label{disc0848}

A photon cluster is found at the angular separation of 12\farcm2 from the gamma-ray
source 3FGL~J0849.3$+$0458, which is in the 2FGL and 2LAC catalogues and is reported as the 
counterpart to the BL Lac object 5BZB~J0849$+$0455.
This source is also in the D$^3$PO catalogue, while there is no high-energy MST detection at these
coordinates. 
However, at energies higher than 10 GeV, this angular separation between the centroid coordinates 
and the blazar results quite higher than the mean association distance (see Sect. 3 and Fig. 1)
and therefore the connection between the cluster and the BL Lac object is rather weak. 
The cluster centroid is at a separation of 5\farcm5 from another BL Lac object, 5BZB~J0848$+$0506, 
that is not yet detected at $\gamma$-ray energies.
The map in Figure~\ref{BZB_due} (right panel) illustrates this situation: a high concentration 
photon cluster is located very close to 5BZB~J0848$+$0506 and not at the position of 
5BZB~J0849$+$0455.
A significant cluster ($M=33$) is also found by MST at energies higher than 30 GeV, where the
mean spatial density of photons in the region is very low. 
On the other hand, when repeating the MST analysis in the LAT field above 3 GeV we find a
unique rich cluster with a very high MST magnitude ($M=88$) and very precisely located at
the 5BZB~J0849$+$0455 coordinates.

In these conditions it is hard to obtain a safe estimate the actual significance of the cluster 
MST~0848$+$0503 and its association with 5BZB~J0848$+$0506 because there is no way to exclude 
photons from the other close bright BL Lac object that can accidentally be found in the MST cluster, 
the richest in the present sample.
Note also that in Table 1 this BL Lac has the highest value of $\mathrm{Log}~ \Phi_{XR}$, while no X-ray
detection of 5BZB~J0849$+$0455 is reported in the 5BZCAT.
One can infer that the former source is a HBL object, and consequently with a particularly hard
$\gamma$-ray spectrum, while the latter might be an LBL object with a possible high energy cut-off.  
In the 3FGL catalogue there are 14 sources associated with BL Lac objects with a radio flux density
at 1.4 GHz lower than 10 mJy but detected in the X rays, as expected for HBL objects.
Therefore, our association is not in contrast with previous results on $\gamma$-ray emission
from this class of sources.

We can conclude, therefore, that both BL Lac objects are high energy emitters but their spectra
are very different.

\subsection{5BZB~J0952$+$7502}
The sky map of the region around 5BZB~J0952$+$7502 has a higher density of photons but
a relative low number of $\gamma$-ray sources (Figure ~\ref{BZB_tre}, left panel).
There are only two MST cluster detections above 10 GeV. 
The first corresponds
to a 3FGL and D$^3$PO source very close to the blazar 5BZU~J1031$+$7441, while the second cluster is 
detected only at energies higher than 10 GeV.
Excluding the farthest photon from the cluster, the new $g$ and $M$ are increased and
also the positional matching improves, making stronger the association with this BL Lac object.
Finally, the spectrum of this source must be rather peculiar with a possible turn off at 
low energies because it is undetected when energies higher than 3 GeV are considered.

\subsection{5BZB~J1311$+$3953}

The detection of the cluster MST~1311$+$3951 is rather critical because it presents three close
photons surrounded by a few other photons which could be background events.
A cluster of 7 photons is found at energies higher than 6 GeV in a $10\degr\times10\degr$ region, 
but with a rather low $g$, whereas there is no detection when a lower energy threshold of 3 GeV
is considered.
This explain the quite low $\sqrt{TS} =2.9$ value found from the ML analysis.
It is interesting to note that the D$^3$PO catalogue reports a source (1DF002707) with a photon 
flux of about 5.0 10$^{-11}$ ph/(cm$^2$ s) and Galactic coordinates $l = 105\fdg811$, 
$b = 76\fdg446$.
The angular distance to the centroid of MST~1311$+$3951 is 8\farcm7, too large for a firm association,
but small enough for suggesting a region with a possible extended emission or source confusion. 

At energies higher than 10 GeV in a 6\degr\ size region MST analysis confirmed this cluster 
($M = 19.48$) and found only one very significant cluster at the position of 5BZB J1309+4305 
(3FGL J1309.3+4304).
In a wider 10\degr\ size region there are three other blazars listed in the Roma-BZCAT, two of them 
associated with 3FGL sources: 5BZB J1305+3855, 5BZQ J1308+3546 (3FGL J1308.7+3545) and 
5BZB J1309+4305 (3FGL J1309.3+4304).
We found at energies higher than 3 GeV again only one very significant cluster corresponding to 
the last one, in any case these blazars are well separated from MST~1311$+$3951 to 
produce any confusion. 

We can conclude that there are some interesting indications for an high energy emission related to 
5BZB~J1311$+$3953, although it is quite difficult to obtain an entirely satisfactory statistical 
significance.

\subsection{5BZB~J1423$+$3737 }\label{s:flaring}

This source is the brightest in the sample and above 3 GeV its position correspond to a high density 
and rich cluster (Figure~\ref{BZB_quattro}, left panel)
In the same field there are two 3FGL  sources (3FGL J1411.1$+$3717 and 3FGL J1419.8$+$3819) at 
a separation higher than 1\degr\ and two D$^3$PO sources, one of which corresponding to the latter 
object and the other close to 5BZG~J1424$+$3705, but none of them associated with this BL Lac object.
Note that this object was already reported as a possible counterpart to 
3EG~J1424$+$3734 in the Third EGRET catalogue \citep{hartman99}, although at a distance 
of about 21\arcmin. It was also present in 1FGL and 1LAC \citep{abdo10b} catalogues, 
but disappeared in the subsequent versions.
The EGRET source was also associated by \cite{colafrancesco02} with the nearby galaxy clusters 
A~1902 and A~1914, whose locations are marked in the map of Figure~\ref{BZB_quattro} by two  
gray circles.

Two light curves of the events at energies above 0.1 and 3 GeV from a circular 
region of radius 0\fdg3 centered at the source position (see Figure~\ref{BZB_quattro}, 
right panel) were extracted.
A flare in which the count rate increased by a factor of about 5, reaching $>$100~MeV fluxes 
around 10$^{-7}$ photons cm$^{-2}$ s$^{-1}$ ($\sim$3$\cdot$10$^{-9}$ photons cm$^{-2}$ s$^{-1}$ 
above 3~GeV) and having a duration of $\sim$3 months is clearly apparent in the period from 
MJD 56550  to 56650 (2013 September--December), making more robust the identification of this 
5BZB~J1423$+$3737 as the counterpart to the $\gamma$-ray source. 
Moreover, this variability and the finding that no high energy photons are observed from the two 
near galaxy clusters seems to discount the possibility that these objects are powerful emitters of  
high energy $\gamma$-rays.

\subsection{5BZB~J2211$-$0003}

This is one of the most crowded field of blazars containing four BL Lac objects and two other type 
blazars (possibly FSRQ, Figure~\ref{BZB_tre}, right panel)
There is only one 3FGL source which corresponds to the unique D$^3$PO source and is associated with 
5BZB~J2206$-$0031. 
MST analysis above 3 GeV gives three significant clusters (open turquoise circles): one corresponding 
to this source, another related to a high concentration of a small number of photons at $l = 62.77$, 
$b = -49.93$, and the third one very close to 5BZB~J2211$-$0003.
The last cluster is the only significant one found at energies higher than 10 GeV.

About the unassociated source we found an interesting possible radio counterpart, NVSS~J223510$-$033332,
at an angular separation of 6.75 \arcmin with a radio flux density at 1.4 GHz of 93 mJy.
It corresponds in SDSS DR10 to a starlike object with $r = 19.45$ mag and $u - r= 0.90$, typical
of many HBL objects detected in the $\gamma$-ray band \citep{massaro12}.

\begin{figure}[h]
\centering
\includegraphics[height=8cm]{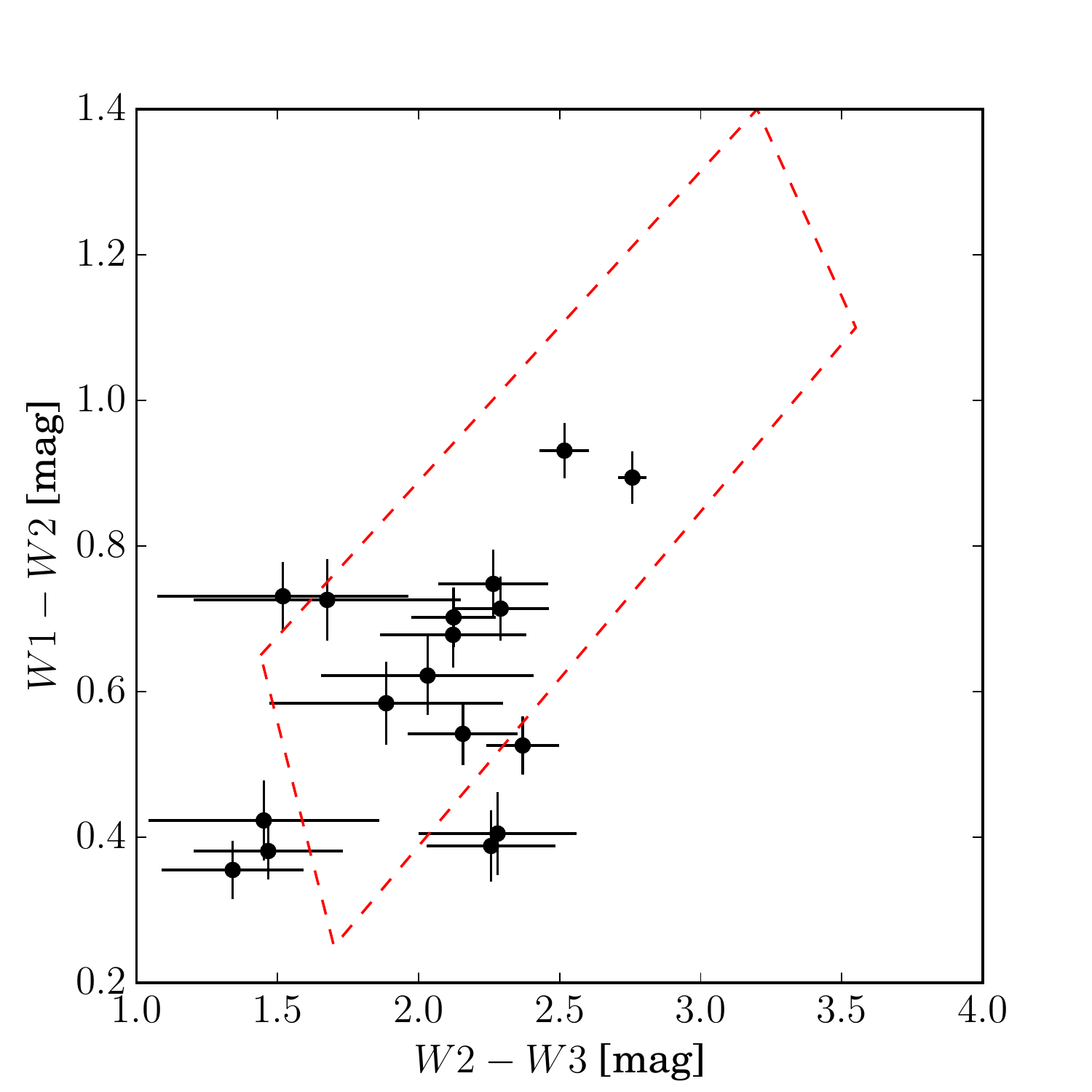}
\caption{Plot of the infrared colours of the MST-selected BL Lac objects from the AllWISE catalogue.
The dashed line define the region where the WISE Blazar strip is located according to the
recent analysis by \cite{dabrusco14}.
The majority of our sources lie within the strip and those outside the boudaries are generally
affected by large errors making possible their location inside the $\gamma$-ray loud region.}
\label{wisecolplot}
\end{figure}

\section{Discussion}

The new detections of $\gamma$-ray clusters closely associated with BL Lac objects at energies 
higher than 10 GeV confirm the high sensitivity of MST method in extracting small photon concentrations 
from a sparse background.
This fact makes the method particularly able in detecting weak sources that, when observed at 
lower energies, could be confused with the background. 
We are confident that our detections are robust, not only for their high $M$ values but also 
because the chance probabilities for such a good positional correspondence are indeed quite low.
The two sources 5BZB~J0801$+$6444 and 5BZB J1216$-$0243 are reported in the 5BZCAT as BL Lac
candidates because their optical spectra are unpublished.
Clearly our detection support their identification as genuine blazars.

We obtained also the significance of these clusters by evaluating their test statistics applying the standard
LAT maximum likelihood method at energies higher than 3 GeV in ROI of 10\degr\ radius: for 14 clusters we 
obtained values of $TS$ higher than 16 (8 of which with $TS > 25$).
For only 5BZB~J0848$+$0506 we failed to obtain convergence because of the proximity of 
5BZB~J0849$+$0455.
These results provide a practical validation of the minimal spanning tree clustering 
method to select candidate sources.

We also verified if the infrared colors of the reported BL~Lac objects are compatible with the WISE
blazar strip, discovered by \cite{massaro11, massarof12} where the majority of $\gamma$-ray loud 
sources of this type are located.
We obtained three band infrared photometric data from the AllWISE catalogue \citep[data at 22~$\mu$m are 
available only for three sources]{cutri14} and computed the $W1 - W2$ and $W2 - W3$ 
colours without reddening correction because all the sources have a Galactic latitude higher than 
25\degr.
Figure~\ref{wisecolplot} shows that the majority of these newly observed BL Lacs above 10 GeV are 
really aligned along the strip, whose approximate countour is marked by the red dashed line on the 
basis of the most recent plots given by \cite{dabrusco14} (WIBRaLS catalogue).
Sources outside the strip have photometric uncertainties large enough that they are also compatible
generally within one standard deviation.
Note, however, that only 5BZB~1437$+$3737 is reported in the WIBRaLS catalogue as class C object because 
it is one of the three sources detected in the 22~$\mu$m WISE bandpass.


\begin{acknowledgements}
We acknowledge use of archival Fermi data. We made large use of the online version of the Roma-BZCAT 
and of the scientific tools developed at the ASI Science Data Center (ASDC),
 of the Sloan Digital Sky Survey (SDSS) archive, of the NED database and other astronomical 
catalogues distributed in digital form (Vizier and Simbad) at Centre de Dates astronomiques de 
Strasbourg (CDS) at the Louis Pasteur University.
\end{acknowledgements}

\bibliographystyle{spr-mp-nameyear-cnd}
\bibliography{mst_bzb.bib} 

\end{document}